\documentclass[twocolumn]{physcon15_nolabel}

\usepackage{graphicx,array}
\usepackage{amssymb,amsmath}

\title{UNI-DIRECTIONAL SYNCHRONIZATION AND FREQUENCY DIFFERENCE LOCKING INDUCED BY A HETEROCLINIC RATCHET}

\author{ \twlsfb \"Ozkan Karabacak
\affiliation{
      Electronics and Communication Engineering Department\\
      Istanbul Technical University\\
      Turkey\\
      karabacak@itu.edu.tr\\
      ozkan2917@gmail.com
    }
   
}

\begin{document}

\maketitle

\begin{abstract}
A system of four coupled oscillators that 
exhibits unusual synchronization phenomena has been analyzed. Existence of a one-way heteroclinic network, called heteroclinic 
ratchet, gives rise to uni-directional (de)synchronization between certain groups of cells. Moreover, we show that locking in
frequency differences occur when a small white noise is added to the dynamics of oscillators.
\end{abstract}


\section{Introduction}

Phase oscillators are used as approximations for the phase dynamics of coupled limit cycle oscillators in the case of weak coupling \cite{kuramoto,pikovsky_01,weakly}. They exhibit synchronization and clustering phenomena \cite{kuramoto,sakaguchi_kuramoto_86}, even if coupling function consist of the first harmonic only. If the second harmonic of the coupling function is considered, it is possible to observe switchings between different clusterings as a result of an asymptotically stable robust heteroclinic cycle \cite{hansel_mato_meunier_93}. It is known that heteroclinic cycles are not structurally stable but they may exist robustly for coupled systems. This is due to the existence of robust invariant subspaces for certain coupling structures that may support robust heteroclinic connections that are saddle-to-sink on the invariant subspaces and form a heteroclinic cycles \cite{krupa_97,ashwin_field_99,aguiar_ashwin_dias_field_11}. Existence of robust heteroclinic cycles or more generally heteroclinic 
networks in a system of three and four globally coupled phase oscillators have been analyzed in \cite{ashwin_burylko_maistrenko_08} and in \cite{ashwin_burylko_maistrenko_popovych_06}, where an extreme sensitivity phenomenon to detuning of natural frequencies has been observed. Namely, oscillators loose synchrony even for very small detuning of natural frequencies. \cite{karabacak_ashwin_10} have considered the third harmonic of the coupling function and observed one-way heteroclinic networks, which are called heteroclinic ratchets. A heteroclinic ratchet is a heteroclinic network that, for some axis, contains trajectories winding in one direction only. Heteroclinic ratchets give rise to extreme sensitivity to detuning of certain sign. Namely, synchronization of a pair of oscillators is possible only when the natural frequency of a certain oscillator is larger than the other. We call this phenomenon uni-directional (de)synchronization.

In the sequel, we identify a heteroclinic ratchet for a system of four coupled oscillators in Section~\ref{sec:model}. Although the system is less complicated than the original ratcheting system considered in \cite{karabacak_ashwin_10}, it exhibits more complicated dynamics: uni-directional synchronization between groups of oscillators, explained in Sections \ref{sec:unidirectional} and \ref{sec:locking}.
%and locking in frequency differences induced by noise and/or detuning.

%In Section~\ref{sec:model}, we introduce the model, find its robust invariant subspaces and identify a robust heteroclinic ratchet supported on these invariant subspaces.

\section{A Model of Four Coupled Oscillators That Supports Heteroclinic Networks}
\label{sec:model}
Consider the following well-known model of coupled phase oscillators:
\begin{equation}
 \dot\theta_i=\omega_i+\frac{K}{N}\sum_{j=1}^{N}c_{ij}g(\theta_i-\theta_j).
\label{sys:oscillator}
\end{equation} 
Here, $\theta_i\in\mathbb T=[0,2\pi)$ denotes the phase of oscillator $i$ and $\omega_i$ is its natural frequency. The connection matrix $\{c_{ij}\}$ represents the coupling. $c_{ij}=1$ if oscillator $i$ receives an input from
oscillator $j$ and $c_{ij}=0$ otherwise. 

Since $g(\cdot)$ is a
$2\pi$-periodic function, it can be written as a sum of Fourier harmonics: $g(x)=\sum_{k=1}^{\infty} r_k\sin (kx+\alpha_k)$. Without loss of generality, we may set $K=N$ and $r_1=-1$ by a scaling of time. Let us choose the following coupling function with two harmonics only:

\begin{equation}
 \label{eq:couplingfunction}
  g(x)=-\sin(x+\alpha_1)+r_2\sin(2x)
\end{equation}

The model (\ref{sys:oscillator}) is used as the approximate phase dynamics of weakly coupled limit cycle oscillators, and the weak coupling gives rise to a $\mathbb T^1$ phase-shift symmetry in the phase model (\ref{sys:oscillator}). Hence, the dynamics of (\ref{sys:oscillator}) is invariant under the phase shift 
$$
(\theta_1,\theta_2,\dots,\theta_N)\mapsto (\theta_1+\epsilon,\theta_2+\epsilon,\dots,\theta_N+\epsilon)
$$
for any $\epsilon\in \mathbb T$. Below, this symmetry is used to reduce the dynamics to an ($N-1$)-dimensional phase difference system. 

\begin{figure}[t]
\begin{center}
\includegraphics{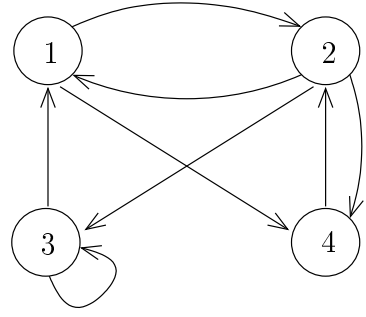}
\end{center}
\caption{
An asymmetric coupled cell system: this gives the coupled system of form (\ref{sys:4dim}).}
\label{fig4cell}
\end{figure}

Let us consider the coupled phase oscillator system (\ref{sys:oscillator}) with the coupling structure given in Figure~\ref{fig4cell}. This gives rise to the following system:

\begin{equation}
\begin{split}
\dot\theta_1&=\omega_1+g(\theta_1-\theta_2)+g(\theta_1-\theta_3)\\
\dot\theta_2&=\omega_2+g(\theta_2-\theta_1)+g(\theta_2-\theta_4)\\
\dot\theta_3&=\omega_3+g(\theta_3-\theta_2)+g(0)\\
\dot\theta_4&=\omega_4+g(\theta_4-\theta_1)+g(\theta_4-\theta_2).
\end{split}
 \label{sys:4dim}
\end{equation}

Defining phase difference variables as $\phi_1:=\theta_1-\theta_2$, $\phi_2:=\theta_2-\theta_4$ and $\phi_3=\theta_3-\theta_4$, we obtain the following dynamical system for phase differences:

\begin{equation}
\begin{split}
\dot{\phi_1}&=\omega_{13}+g(\phi_1+\phi_3-\phi_2)+g(\phi_1)\\ 
& \quad \quad -g(\phi_3-\phi_2)-g(0)\\
\dot{\phi_2}&=\omega_{24}+g(-\phi_1-\phi_3+\phi_2)+g(\phi_2)\\
& \quad \quad -g(-\phi_3-\phi_1)-g(-\phi_2)\\
\dot{\phi_3}&=\omega_{34}+g(\phi_3-\phi_2)+g(0)\\
& \quad \quad -g(-\phi_3-\phi_1)-g(-\phi_2).
\end{split}
\label{sys:3dim}
\end{equation}

$\omega_{ij}$ denotes the detuning between oscillator $i$ and oscillator $j$, namely
 $\omega_{ij}=\omega_i-\omega_j$.

We first assume identical oscillators, that is 
\begin{equation}
\label{eq_identity}
	\omega_1=\cdots=\omega_4=\omega \Longrightarrow \omega_{ij}=0 \quad \forall i,j.
\end{equation}
Oscillators with different natural frequencies will be considered in Section~\ref{sec:unidirectional}. %and \ref{sec:locking}.

\subsection{Invariant Subspaces}
\label{sec:invariant}
The assumption that the oscillators are identical makes it possible to use the balanced coloring method \cite{stewart_golubitsky_pivato_03} to obtain invariant subspaces of the system~(\ref{sys:4dim}). A coloring of cells in a coupled cell system is called balanced if cells with identical color receives the same number of inputs from cells of any given color. A balanced coloring gives rise to an invariant subspace obtained by assuming that the cells of same color have identical states. The converse of this statement is also true. Namely, for a given coupling structure, a coloring is balanced if the corresponding subspace is invariant under any system having that coupling structure. Therefore, the invariant subspaces obtained by the balanced coloring method are robust under the perturbations that preserve the coupling structure. For an introduction to this theory, see \cite{golubitsky_stewart_06}.

\begin{figure}[t]
\begin{center}
\includegraphics[width=0.45\textwidth]{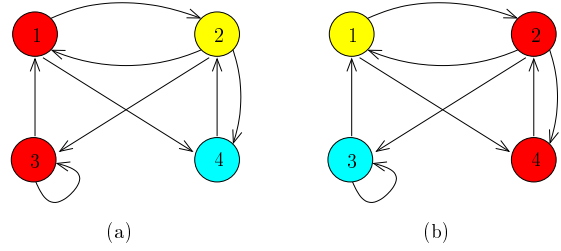}
\end{center}
\caption{
Two balanced colorings of the coupled cell system in Figure~\ref{fig4cell}. These give rise to 2-dimensional invariant subspaces $X_1$ and $X_2$ of the system~(\ref{sys:3dim}).}
\label{fig:type1_colors}
\end{figure}

Using the balanced coloring method the invariant subspaces of the coupled cell system given in Figure~\ref{fig4cell} can be found as in Table~\ref{tab:invariant}. Using the above-mentioned phase difference reduction, the corresponding invariant subspace in $\mathbb T^3$ for the system (\ref{sys:3dim}) are also listed in Table~\ref{tab:invariant}. Note that for the system (\ref{sys:3dim}), there are only two 2-dimensional invariant subspaces, namely $X_1$ and $X_2$. Balanced colorings for these invariant subspaces are given in Figure~\ref{fig:type1_colors}. The invariant subspaces $X_1$ , $X_2$ and their intersection $X_3$ can support a robust heteroclinic cycle (see \cite{ashwin_karabacak_nowotny_11} for robustness criteria of heteroclinic cycles). 
\begin{table}
\centering
	\begin{small}\begin{tabular}{||l|l|l||}
		\hline\hline
		Balanced & Invariant subspaces of the system (\ref{sys:4dim})\\
		colorings & on $\mathbb T^4$ and system (\ref{sys:3dim}) and on $\mathbb T^3$\\
		\hline\hline
		$\{1|2|3|4\}$ & $X_0=\mathbb T^4$ \\ & $\bar X_0=\mathbb T^3$ (whole space) \\ \hline
		$\{1|2|34\}$ & $X_1=\{\theta\in\mathbb T^4\mid\theta_2=\theta_4\}$ \\ & $\bar X_1=\{\phi\in\mathbb T^3\mid\phi_2=0\}$ ($\phi_1-\phi_3$ plane)\\ \hline
		$\{12|3|4\}$ & $X_2=\{\theta\in\mathbb T^4\mid\theta_1=\theta_3\}$ \\ & $\bar X_2=\{\phi\in\mathbb T^3\phi_1=0\}$ ($\phi_2-\phi_3$ plane)\\ \hline
		$\{12|34\}$ & $X_3=\{\theta\in\mathbb T^4\mid\theta_1=\theta_3,\theta_2=\theta_4\}$ \\ & $\bar X_3=\{\phi\in\mathbb T^3\phi_1=\phi_2=0\}$ ($\phi_3$ axis)\\ \hline
		$\{134|2\}$ & $X_4=\{\theta\in\mathbb T^4\mid\theta_1=\theta_3=\theta_4\}$ \\ & $\bar X_4=\{\phi\in\mathbb T^3\phi_1=\phi_3=0\}$ ($\phi_2$ -axis)\\ \hline 
		$\{1234\}$  & $X_5=\{\theta\in\mathbb T^4\mid\theta_1=\theta_2=\theta_3=\theta_4\}$ \\ & $\bar X_5=\{(0,0,0)\}$ (origin) \\ \hline
		\hline
	\end{tabular}	\end{small}
\caption{Balanced colorings of the coupled system given in Figure \ref{fig4cell}. The corresponding invariant subspaces for the system (\ref{sys:4dim}) and the corresponding reduced invariant subspaces for the system (\ref{sys:3dim}) are given.}
\label{tab:invariant}
\end{table}

\subsection{Existence and Stability of a Heteroclinic Ratchet}

A heteroclinic ratchet (first defined in \cite{karabacak_ashwin_10}) is a heteroclinic network that contains a heteroclinic cycle winding in some direction and does not contain another heteroclinic cycle winding in the opposite direction. To be precise, a heteroclinic cycle $C\subset\mathbb T^N$ parametrized by $x(s)$ $(x\colon[0,1)\rightarrow\mathbb T^N)$ is \emph{winding in some direction} {if there is a projection map $P\colon\mathbb R^N\rightarrow\mathbb R$ such that the parametrization $\bar x(s)$ $(\bar x\colon[0,1)\rightarrow\mathbb R^N)$ of the lifted heteroclinic cycle $\bar C\subset\mathbb R^N$ satisfies $\lim_{s\rightarrow 1}P(\bar x(s))-P(\bar x(0))=2k\pi$} for some positive integer $k$. A heteroclinic cycle winding in the opposite direction would satisfy the same condition for a negative integer $k$ (see \cite{ashwin_karabacak_11} for general properties of heteroclinic ratchets).

As discussed above, the system (\ref{sys:3dim}) may have a robust heteroclinic network on the invariant subspaces $X_1$ and $X_2$. Such a heteroclinic network should be connecting saddles on $X_3=X_1\cap X_2$. Reducing the equations in (\ref{sys:3dim}) to $X_3$ and considering identical natural frequencies, we get
\begin{equation}
 \label{sys:1dim}
 \dot\phi_3=g(\phi_3)-g(-\phi_3).
\end{equation}
This system is $\mathbb Z_2$-equivariant, and therefore it can admit a codimension-1 pitchfork bifurcation of the zero solution under some nondegeneracy conditions. The saddles emanating from this bifurcation are on $X_3$ and they are of the form $p=(0,0,p_3)$ and $q=(0,0-p_3)$. The value $p_3$ can be obtained by solving $g(p_3)-g(-p_3)=0$ as 
\begin{equation}
 p_3=\cos^{-1}\left(\frac{\cos\alpha_1}{2r_2\cos\alpha_2}\right).
\label{eq:p3}
\end{equation}

%\begin{proposition}
% The following conditions are necessary for the existence of a robust heteroclinic cycle on $X_1\cup X_2$ for the system~\ref{sys:3dim}:
%\end{proposition}

\begin{figure}[t]
\begin{center}
\includegraphics[width=0.45\textwidth]{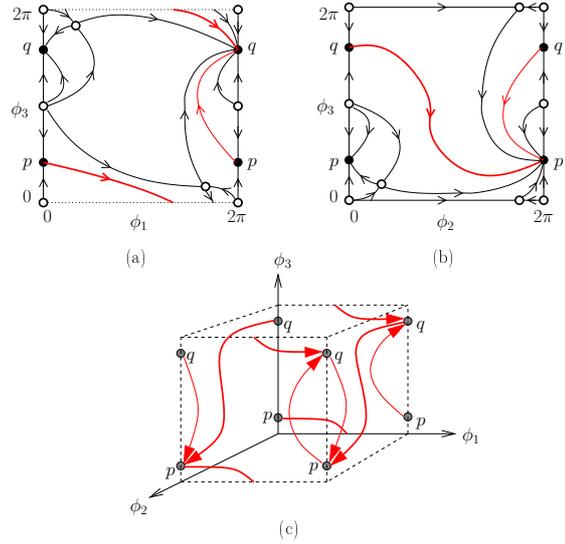}
\end{center}
\caption{
Phase portraits of the system~(\ref{sys:3dim}) on invariant subspaces $\bar X_1$ (a) and $\bar X_2$ (b) are illustrated for parameters given in  (\ref{eq:parameterset}). Red lines indicate robust heteroclinic trajectories. Thick red lines are the winding heteroclinic trajectories. The robust heteroclinic ratchet formed by these winding and non-winding heteroclinic trajectories and the saddles $p$ and $q$ is shown in (c).}
\label{fig:allinone}
\end{figure}

In order to show that there exist heteroclinic connections between $p$ and $q$ we use the simulation program XPPAUT \cite{ermentrout}. We identify a heteroclinic ratchet for the parameter values 
\begin{equation}
 \label{eq:parameterset}
\alpha_1=-5, \quad r_2=0.15
\end{equation}
(see Figure~\ref{fig:allinone}). On $X_1$, the heteroclinic ratchet contains a non-winding trajectory and a trajectory winding along $+\phi_1$ and $-\phi_3$ directions (see Figure~\ref{fig:allinone}a). On $X_2$, it contains a non-winding trajectory and a trajectory winding along $+\phi_2$ direction  (see Figure~\ref{fig:allinone}b). These four connections and the saddles $p$ and $q$ form a heteroclinic ratchet (see Figure~\ref{fig:allinone}c). For the parameters given in (\ref{eq:parameterset}), $p$ can be found as $(0,0,0.3315)$.  
Considering the Jakobien of (\ref{sys:3dim}) at $p$, we can find the eigenvalues of the saddle $p$ as $\lambda^{(p)}_1=-0.3112$, $\lambda^{(p)}_2=0.2967$ and $\lambda^{(p)}_3=-0.0636$. Similarly, the eigenvalues of $q=-p=(0,0,0.3315)$ can be found as $\lambda^{(q)}_1=0.3130$ and $\lambda^{(q)}_2=-0.3276$ and $\lambda^{(q)}_3=-0.0636$. These eigenvalues of $p$ and $q$ correspond to the eigenvectors $v_1=(1,0,0)\in X_1$, $v_2=(0,1,0)\in X_2$ and $v_3=(0,0,1)\in X_3$. A heteroclinic cycle is attracting if the saddle quantity, defined as the absolute value of the ratio between the product of the eigenvalues corresponding to the expanding connections and the product of the eigenvalues corresponding to the contracting connections is smaller than 1 \cite{melbourne_89}. Hence, we can conclude that the heteroclinic ratchet for the system (\ref{sys:3dim}) with parameters given in (\ref{eq:parameterset}) is asymptotically stable since the saddle quantity $\left|\frac{\lambda_2^{(p)}\lambda_1^{(q)}}{\lambda_1^{(q)}\
lambda_2^{(p)}}\right|=0.9532$ is less than one.  

\begin{figure}[t]
\hspace{-0.5cm}
\includegraphics[width=0.4\textwidth, angle=-90]{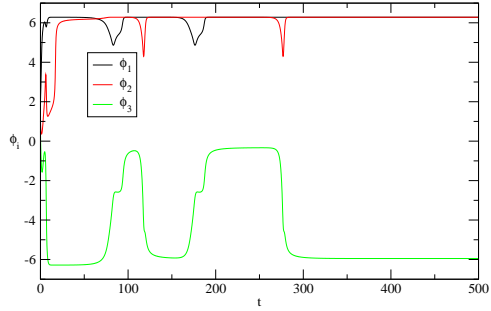}
\vspace{-1.5cm}
\caption{
A solution of the system~(\ref{sys:3dim}) converging to the heteroclinic ratchet. Initial states are chosen as $(2,1,0.5)$. The solution shows the peculiar property for heteroclinic cycles that the residence time near equilibria increases as $t\to\infty$, before $\phi_1$ and $\phi_2$ get locked at zero due to the precision errors.}
%\vspace{-1.2cm}
\label{fig:converging}
\end{figure}

A solution of (\ref{sys:3dim}) converging to the heteroclinic ratchet can be seen in Figure~\ref{fig:converging}. 
The increase in the residence time near equilibria is typical for a solution converging to a heteroclinic network. 
Winding of $\phi_1$ and $\phi_3$ occur at the same time, respectively in positive and negative directions, due to the winding heteroclinic trajectory on $X_1$ (see Figure~\ref{fig:allinone}a). 
Winding of $\phi_2$ occur in the positive direction due to the winding heteroclinic trajectory on $X_2$ (see Figure~\ref{fig:allinone}b). Since at each turn the solution gets closer to the equilibria $p$ and $q$, after some time, $\phi_1$ and $\phi_2$ get locked at zero due to the precision errors. 
Note that, the invariant subspaces $X_1$ and $X_2$ serve as barriers, and therefore no solution can pass through them. 
For this reason the solution winds in $\phi_1$ and $\phi_3$ directions only one time. Since winding in $-\phi_3$ direction occurs together with the winding in $+\phi_1$ direction, this also happens only one time. 
However, these barriers 
can be broken by noise and/or detuning of natural frequencies leading to the uni-directional synchronization phenomenon.

\section{Uni-directional Synchronization in the Model}
\label{sec:unidirectional}

\begin{figure}[t]
%\begin{center}
\hspace{-0.5cm}
\includegraphics[width=0.4\textwidth, angle=-90]{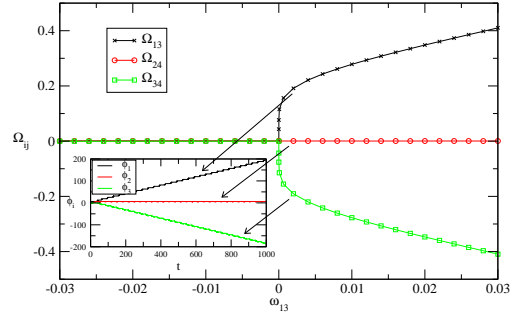}
%\end{center}
\vspace{-1.5cm}
\caption{Uni-directional synchronization with respect to $\omega_{13}$. The main graph shows the frequency differences $\Omega_{13}$, $\Omega_{24}$ and $\Omega_{34}$ for (\ref{sys:4dim}) with parameters given in (\ref{eq:parameterset}) as a function of detuning $\omega_{13}$ when $\omega_{24}=\omega_{34}=0$.  Oscillators are frequency synchronized when $\omega_{13}\leq 0$ and the synchronization fails for oscillator pairs $(13)$ and $(24)$ whenever $\omega_{13}>0$. The insets show time evolution of the phase differences $\phi_i$ for a positive value of $\omega_{13}$.}
\label{fig:detuning1}
\end{figure}

We say that oscillators $i$ and $j$ are (frequency) synchronized if the observed frequency differences 
\begin{equation}
\label{eq:Omega}
\Omega_{ij}=\lim_{t\to\infty}\frac{|\theta^{L}_i-\theta^{L}_j|}{t},
\end{equation}
is equal to zero. Here $\theta^{L}_i\in\mathbb R$ is the lifted phase variable for $\theta_i\in\mathbb T$.
It is know that coupled oscillators can get frequency synchronized when the distance between their natural frequencies, namely $|\omega_{ij}|:=|\omega_i-\omega_j|$ is small enough. If frequency synchronization of oscillators $i$ and $j$ occurs only when a specific one of the oscillators has greater natural frequency, namely for a certain sign of $\omega_{ij}$, we say that synchronization is uni-directional. Uni-directional synchronization phenomenon has been shown to occur for oscillator pairs when an asymptotically stable heteroclinic ratchet exists in the phase space \cite{karabacak_ashwin_10,ashwin_karabacak_11}. 

\begin{figure}
%\begin{center}
\hspace{-0.5cm}
\includegraphics[width=0.4\textwidth, angle=-90]{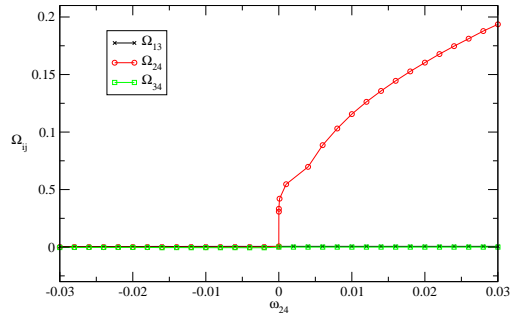}
\vspace{-1.5cm}
%\end{center}
\caption{Uni-directional synchronization with respect to $\omega_{24}$. The frequency differences $\Omega_{13}$, $\Omega_{24}$ and $\Omega_{34}$   for ({\protect
\ref{sys:4dim}})  with parameters given in (\ref{eq:parameterset}) are shown as a function of detuning $\omega_{24}$ when $\omega_{13}=\omega_{34}=0$.  Oscillators are frequency synchronized when $\omega_{24}\leq 0$ and the synchronization fails for the oscillator pair $(24)$ whenever $\omega_{24}>0$.}
\label{fig:detuning2}

\end{figure}
\begin{figure}[t]
%\begin{center}
\hspace{-0.5cm}
\includegraphics[width=0.4\textwidth, angle=-90]{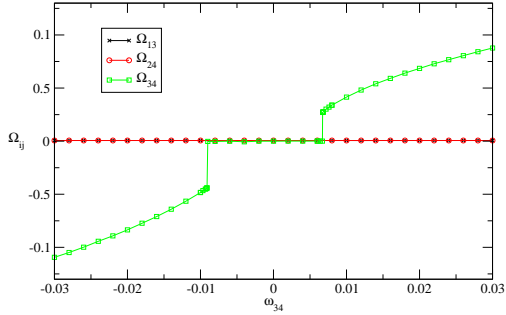}
%\end{center}
\vspace{-1.5cm}
\caption{Bi-directional synchronization (the usual case) with respect to $\omega_{34}$. The frequency differences $\Omega_{13}$, $\Omega_{24}$ and $\Omega_{34}$   for (\ref{sys:4dim}) with parameters given in (\ref{eq:parameterset}) are shown as a function of detuning $\omega_{34}$ when $\omega_{13}=\omega_{24}=0$.  Oscillators are frequency synchronized when $|\omega_{34}|$ is small enough and the synchronization fails for the oscillator pair $(34)$ for large values of $|\omega_{34}|$.}
\label{fig:detuning3}
%\vspace{-0.3cm}
\end{figure}

For the system (\ref{sys:3dim}), we investigate the effect of detunings $\omega_{13}$, $\omega_{24}$ and $\omega_{34}$ on the synchronization of oscillators, respectively in Figure~\ref{fig:detuning1}, \ref{fig:detuning2} and \ref{fig:detuning3}. Due to the winding connections in the heteroclinic ratchet, uni-directional synchronization occurs for detunings $\omega_{13}$
 and $\omega_{24}$. However, because of the connection winding both in $+\phi_1$ and $-\phi_3$ directions, a positive detuning $\omega_{13}$ leads to synchronization of oscillators $1,2$ and $4$. This is because $\theta_1-\theta_3\cong -(\theta_3-\theta_4)\Longrightarrow \theta_1\cong \theta_4$. The synchronized groups of oscillators for each detuning case are given in Table~\ref{tab:detunings}. It is interesting that the oscillators $1,2$ and $4$ get synchronized for a positive detuning of $\omega_{13}$, although the space $\{\theta\in\mathbb T^4\mid \theta_1=\theta_2=\theta_4\}$ is not one of the synchronization spaces obtained by the balanced coloring method in Section~\ref{sec:invariant}. Hence, it is not an invariant subspace.  

\begin{table}
\centering
\begin{footnotesize}
	\begin{tabular}{||l|l|l|l||}
		\hline\hline
		Detuning&Natural&Winding&Observed\\
		Direction&Frequencies&Direction&Frequencies\\
		\hline\hline
		$+\omega_{13}$&$\omega^+,\omega,\omega,\omega$&$+\phi_1,-\phi_3$&$\Omega^+,\Omega^+,\Omega,\Omega^+$\\
		$+\omega_{24}$&$\omega,\omega^+,\omega,\omega$&$+\phi_2$&$\Omega,\Omega^+,\Omega,\Omega$\\
		$+\omega_{34}$&$\omega^+,\omega,\omega^+,\omega$&0&$\Omega,\Omega,\Omega,\Omega$\\
		\hline
	\end{tabular}\end{footnotesize}
\caption{The effect of detunings on the synchronization of oscillators for the system (\ref{sys:3dim}). These can be obtained from Figures~\ref{fig:detuning1}, \ref{fig:detuning2} and \ref{fig:detuning3}. Negative detunings have not been considered as they have no effect. $\omega^+$ ($\Omega^+$) represents a number slightly larger than $\omega$ ($\Omega$).}
\label{tab:detunings}
\end{table}           

\section{Locking in Frequency Differences}
\label{sec:locking}

\begin{figure}[t]
%\begin{center}
\hspace{-0.5cm}
\includegraphics[width=0.4\textwidth, angle=-90]{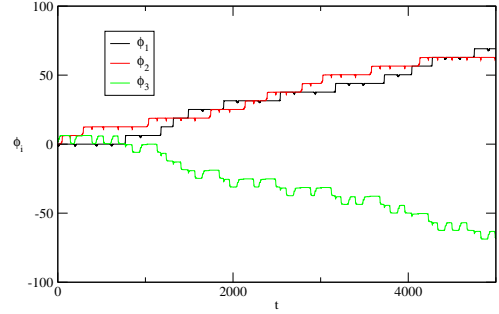}
%\end{center}
\vspace{-1.5cm}
\caption{
A solution of the system (\ref{sys:3dim}) for parameters given in (\ref{eq:parameterset}) with no detuning and with additive white noise (amplitude$=10^{-12}$).}
\label{fig:noise}
\end{figure}

Noise induced uni-directional desynchronization of oscillators has been observed in \cite{karabacak_ashwin_10} for a coupled system admitting a heteroclinic ratchet. Here, we show that existence of a heteroclinic ratchet for the system (\ref{sys:3dim}) leads to a locking in frequency differences when a small noise is applied. Figure~\ref{fig:noise} shows a solution of the system~(\ref{sys:3dim}) under white noise with amplitude $10^{-12}$. The noisy solution exhibits approximately equal number of windings in $\phi_1$ and $\phi_2$ directions. This is because the noise is homogeneous and the invariant subspace $X_1$ (resp. $X_2$) divides any $\epsilon$-ball around the equilibrium $q$ (resp. $p$) into two regions of attractions of equal volume for the winding and non-winding trajectories. On the other hand, the number of windings in $\phi_1$ and $-\phi_3$ directions are exactly the same because of the structure of the heteroclinic ratchet in Figure~\ref{fig:allinone}.

As a result, the solution gives rise to the following frequency locking between frequency differences:
\begin{equation}
 \label{eq:locking}
 \Omega_{13}=\Omega_{24}=-\Omega_{34}.
\end{equation}
This is in agreement with the simulation results given in Figure~\ref{fig:noise}.
Therefore, the observed oscillator frequencies $\Omega_i:=\lim_t|\theta^L_i(t)|/t$  are in the following form: 
\begin{equation}
 \label{eq:locked_osc}
 \left(\begin{array}{l} 
 \Omega_1\\ \Omega_2\\ \Omega_3\\ \Omega_4 \end{array}\right)
 =
 \left(\begin{array}{l} 
 \Omega+\delta\\ \Omega\\ \Omega+\delta\\ \Omega+2\delta \end{array}\right),
\end{equation}
where $\delta$ is a positive number. This type of a result cannot be seen directly from the connection structure of the coupled system, and is a consequence of the particular heteroclinic ratchet that the system admits. Although the noise induces synchronization of oscillators $1$ and $3$, the balanced coloring method explained in Section~\ref{sec:invariant} does not give $\{\theta\in\mathbb T^4\mid\theta_1=\theta_3\}$ as an invariant subspace. 

\section{Conclusion}
We have analyzed a system of four coupled phase oscillators. The existence of an asymptotically stable heteroclinic ratchet gives rise to uni-directional synchronization of certain groups of oscillators and induce a particular locking in the frequency differences of oscillators when small amplitude white noise is introduced to the system. These phenomena also lead to frequency synchronization of some oscillators, that can not be found by using the balanced coloring method, therefore does not correspond to any synchrony subspace.

For future works, the relation between the connection structure and possible synchronization groups can be studied. Although the synchronization groups can not be inferred from the coupling structure directly, the coupling structure serves to create invariant subspaces on which heteroclinic ratchets can be supported. For this reason, the coupling structure plays an indirect role on the existence of possible synchronization groups. Another direction could be to study bifurcations of heteroclinic ratchets that result in winding periodic orbits on torus. This can explain the effect of small detunings of natural frequencies on the observed frequencies in a complete way.

%% If you want to use Bibtex, then remove the thebibliography environment above,
%% and uncomment the \bibliography command below as well as \usepackage[dcucite]{harvard} at the beginning of the document.
%\input{./difference_locking.bbl}
\bibliography{difference_locking}
\bibliographystyle{apalike}

\end{document}